\documentclass{aa}  
\usepackage{graphicx}
\usepackage{amsmath}	
\usepackage{amssymb}	
\usepackage{multicol}        
\usepackage{multirow}
\usepackage{lscape}
\usepackage{longtable} 
\usepackage{xspace}

\usepackage{natbib}
\usepackage{txfonts}
\usepackage{adjustbox}
\usepackage{color}
\usepackage{rotating}
\usepackage{supertabular}
\usepackage[usenames,dvipsnames,svgnames,table]{xcolor}
\usepackage{url}
\usepackage{booktabs}
\usepackage{siunitx}

\newcommand{\sfg}{\mathrm{S^{4}G}}
\newcommand{\micron}{$\mu$m}

\begin{document}

   \title{Infrared-detected AGNs in the Local Universe}


 \author{T. \.{I}kiz\inst{1,2,3}\thanks{\email{tuba.ikiz@atauni.edu.tr}} \and
        R.~F.~Peletier\inst{1} \and
        P.~D.~Barthel\inst{1}
       \and
       C. Ye\c{s}ilyaprak\inst{2,3}
       }
  \institute{Kapteyn Astronomical Institute, University of Groningen, PO Box 800, NL-9700 AV Groningen, the Netherlands
        \and Department of Astronomy and Astrophysics, Faculty of Science, Atat\"{u}rk University, Erzurum, 25240, Turkey
	\and Atat\"{u}rk University Astrophysics Research and Application Center (ATASAM), Erzurum, 25240, Turkey }

   \date{Received May 28, 2019; accepted June 10, 2020}

  \abstract
{\textit{Spitzer}/IRAC color selection is a promising technique to identify hot accreting nuclei, that is to say AGN, in galaxies. We demonstrate this using a small sample of SAURON galaxies, and explore this further.}
{Goal of this study is to find a simple and efficient way to reveal optically obscured nuclear accretion in (nearby) galaxies.}
{We apply an infrared selection method to the \textit{Spitzer Survey of Stellar Structures in Galaxies} ($\sfg$) sample of more than 2500 galaxies, together with its extension sample of more than 400 galaxies. We use the \textit{Spitzer} colors to find galaxies in the S$^{4}$G survey containing a hot core, suggesting the presence of a strong AGN, and study the detection fraction as a function of morphological type. We test this infrared color selection method by examining the radio properties of the galaxies, using the VLA NVSS and FIRST surveys.
}
{Using the radio data, we demonstrate that galaxies displaying hot mid-infrared nuclei stand out as being (candidate) active galaxies. When using, instead of \textit{Spitzer}, colors from the lower spatial resolution \textit{WISE} mission, we reproduce these results. Hence multi-band infrared imaging represents a useful tool to uncover optically obscured nuclear activity in galaxies.}
{}


\maketitle

\section{Introduction}
\label{sec:1}

Galaxies are the basic building blocks of the Universe, and understanding their formation and evolution is crucial to many areas of current astrophysical researches. Nearby galaxies contain the \textquoteleft fossil record\textquoteright\ of the evolution of galaxies in great detail and provide a wealth of detail to comprehensively test existing galaxy formation and evolution models. A galaxy\textquoteright s structure is linked to both its mass and evolutionary history. Although galaxies show a great deal of variety, one thinks that their current structures are principally determined by their masses, environment and available amounts of gas \citep[e.g.][] {Sheth2013}.

Galaxies are known to contain supermassive black holes \citep[e.g.][]{Fr}, whose masses correlate with the masses of their bulges, implying that the processes of forming bulges and supermassive black holes are intimately connected \citep{Magorrian,Tremaine2002}. These supermassive black holes are located in the nuclei of probably all galaxies. A small fraction of galaxies also has a very bright nucleus called an Active Galactic Nucleus (AGN), showing excess emission thought to be due to accretion of mass by the supermassive black hole that exists at the center of the galaxy. It is thought that AGN play a significant role during the formation of galaxies by creating large outflows that quench star formation in the galaxy \citep{kormendy&ho}. The complex central regions of active and inactive galaxies are thus an ideal laboratory for studying the evolution and formation of galaxies. 

AGN can be defined in several ways at essentially all wavelengths (X-rays, mid-IR, and radio detections and optical emission line diagnostics). A number of spectroscopic surveys of nuclei of nearby galaxies have been performed \citep[e.g.][]{Heckman1980,Ho1995} and large surveys of the local universe such as the 2 degree field survey (Colless et al. 2003) and the Sloan Digital Sky Survey (SDSS; \citet{York2000}) have characterized properties of AGNs with an unprecedented statistical accuracy \citep[e.g.][]{kauffmann&heckman,Heckman2004,Kewley2006,Schawinski2007}. Optical spectroscopic surveys of local galaxies reveal that the nuclear activity in nearby galaxies is generally weak and shows a dichotomy with Hubble type. Emission line ratio diagnostics indicate that two-thirds of early-type and early-type spiral galaxies (E-Sb) and only about 15\% of late-type galaxies (Sc-Sd) host some kind of active nuclei (Palomar survey; \citet{Ho2008}). Summed over all Hubble types 40\% of the nearby galaxies can be considered active, on the basis of optical spectroscopy. 

At near-infrared wavelengths (2-5 \micron) AGN also show prominent emission. A good tool to examine this AGN feature are JHKL (1--3.8 \micron) color diagrams, since stellar and non-stellar emission of galaxy nuclei have different locations in these diagrams \citep{Glass1985,Alonso1998}. The infrared color-color diagrams are useful method to realize whether reddening is affecting the colors and to discovering whether an infrared excess may be present. Nearby active nuclei show a near-infrared excess in comparison to normal nuclei which are dominated by stellar and star formation emission \citep{de_grijp,Gallimore2010}. The near-infrared emission of active nuclei can be explained as arising from either hot dust or a power-law (non-thermal) spectral energy distribution. The effects of dust extinction can be partially refined with using the near-IR data, but only up to some limit. Near-infrared surveys that only go to K-band are not effective to diagnose AGN, because of the stellar contamination.

Selection of AGN in the mid-infrared allows the exploration of strong AGN and quasars whose optical and soft X-ray emission is hidden by dust \citep[e.g.][]{Lacy_2004,Lacy_2007,Lacy2015,stern2005,Martinez2005,Stern_2012,Donley_2012,Eisenhardt2012}. Mid-infrared selection has the robust characteristic that it provides an opportunity for the selection of samples of AGNs containing both moderately obscured and unobscured objects of similar bolometric luminosities, allowing an estimate to be made of the importance of the obscured AGN population to the AGN population as a whole \citep{Lacy2015}. In the mid-infrared, the presence and absence of infrared emission features associated with polycyclic aromatic hydrocarbons (PAHs) and high-ionization infrared lines \citep{Laurent} ensure strong AGN diagnostics.

Surveys that could in principle be used to find AGN in galaxies are Infrared Astronomical Satellite [IRAS] \citep{IRAS1984}, Deep Near Infrared Survey of the Southern Sky [DENIS] \citep{Epchtein,Paturel}, Two Micron All Sky Survey [2MASS] \citep{Skrutskie} and more recently the Wide-field Infrared Survey Explorer ([WISE], \citet{Wright2010}. However, the spatial resolution of the IRAS data at 12, 25, 60 and 100 {\micron}, is too low ($\sim$ 1\arcmin\ or lower). [DENIS] and [2MASS] observed the sky at wavelenghts of 2{\micron} and below: most of the emission these surveys detect from galaxies comes from stars, making it difficult to detect light from the AGN. There have not been any comprehensive infrared spectroscopic surveys designed to find nearby active galaxies.

This all changed with the \textit{Spitzer Space Telescope}, and -- at somewhat lower resolution -- the \textit{WISE} mission. The past few years have seen numerous actual studies using \textit{Spitzer} and \textit{WISE} observations to find and study AGN at high redshift. There are various mid-infrared color selections, such as the \textit{Spitzer} two-color criteria in \citet{Lacy_2004,stern2005,Lacy_2007}, and \citet{Donley_2012}, and the \textit{WISE} two-color criteria in \citet{Jarrett_2011} and \citet{Mateos_2012} and the \textit{WISE} one-color criteria of \citet{Stern_2012} and \citet{Assef_2013}. All of these, using different AGN samples, are in rough agreement with each other and distinguish AGN cleanly from stars and star-forming galaxies in the mid-infrared color space. Mid-infrared selection has the robust characteristic that it permits the selection of samples of AGN containing both moderately obscured and unobscured objects of similar bolometric luminosities, allowing an estimate to be made of the importance of the obscured AGN population to the AGN population. It should be noted however that mid-infrared color selection of AGN is far from complete \citep[e.g.][]{radcliffe2019}. 

Radio continuum selection is one of the original ways to find AGN. Whether or not in combination with radio jets, compact, flat spectrum, high brightness temperature radio cores are a defining feature of an accreting galactic nucleus \citep{Mushotzky2004}. At the same time, care must be taken in interpreting low resolution data, as weak extended radio emission can draw from star formation. Most radio data have fairly accurate positions (better than 45{\arcsec} for the largest radio survey, the National Radio Astronomy Observatory (NRAO) Very Large Array (VLA) Sky Survey or NVSS, and better than 7{\arcsec} for the most sensitive large solid angle survey, FIRST), allowing counterparts in other wavelength bands to be readily identified. 

The \textit{Spitzer} IRAC study of \citet{stern2005} showed that for unresolved AGN the [3.6]-[4.5] mid-infrared color was considerably redder than for non-AGN or stars \citep{Eisenhardt2004}. Building further upon \citet{vanderwolk2011}, we here set out to use the nuclear mid-IR [3.6]-[4.5] color selection, that is to say the hot core selection in an attempt to uncover obscured and non-obscured actively accreting central black holes in the large extended galaxies of the $\sfg$ galaxy sample. This \textit{Spitzer} Survey of Stellar Structure in Galaxies, $\sfg$ is specifically designed to answer the need for a deep, large and uniform IR survey of nearby galaxies (see \citet{Sheth} for the full survey description).

This paper is organised as follows: Section~\ref{sec:2} briefly describes the mid-IR hot core selection method applied to a well studied sample of nearby galaxies. Section~\ref{sec:3} describes the $\sfg$ data and sample construction. Section~\ref{sec:4} describes our data analysis; we subsequently deal with the AGN fraction in the $\sfg$ sample, the relation between galaxy luminosity and the presence of an AGN, and confirmation with \textit{WISE} data. Section~\ref{sec:5} presents the discussion and conclusions. In Appendix~\ref{AppA} we present  the \textit{Spitzer} and \textit{WISE} color comparison, followed by lists of the AGN defined from the \textit{Spitzer} central colors in Appendix~\ref{AppB}.
\section{MID-IR DETECTION METHOD APPLIED TO NEARBY GALAXIES}
\label{sec:2}

\begin{figure*}
\centering 
\resizebox{0.70\hsize}{!}{\includegraphics[]{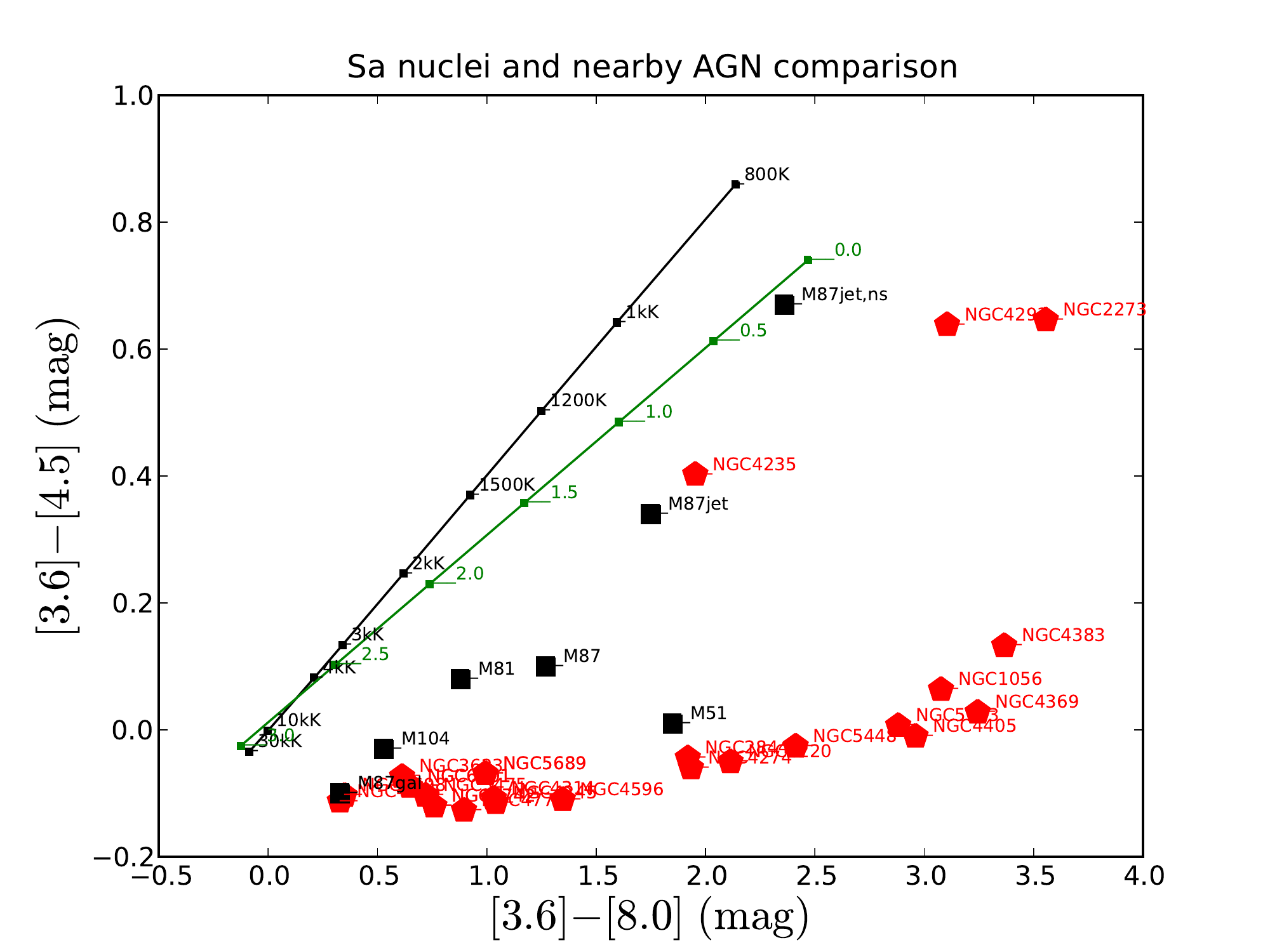}}
\caption{Spitzer color-color diagram. SAURON Sa sample nuclei are plotted in red and nearby Messier AGN in black. For M\,87 we also indicate the colors of the large-scale jet at knots A/B indicated as M87jet, the stellar emission at the same radius as the jet (M87 gal) and the stellar-subtracted jet emission (M87jet, ns). Black bodies at various temperatures are plotted with a black line. Power-law models $\nu F_{\nu} \propto \nu^{\alpha}$ at various spectral indices $0\leq\alpha\leq3$ are plotted with a green line.}
\label{agnind}
\end{figure*}

To illustrate the hot core method, we first show in Fig.~\ref{agnind} the [3.6]-[4.5] vs. [3.6]-[8.0] color-color diagram for the sample of 24 Sa spiral galaxies observed as part of the SAURON survey \citep{deZeeuw2002, Peletier2007, Falcon2006, Shapiro2008}, supplemented with four well-known nearby AGN: M\,51/NGC\,5194 (Seyfert), M\,87/NGC\,4486/3C\,274 (radio-loud LINER), M\,81/NGC\,3031 (Seyfert), and M\,104/NGC\,4594 (Sombrero-galaxy: LINER) -- see Table \ref{tab:nagn}.
It is readily seen that three SAURON spirals stand out: NGC\,2273, NGC\,4235, and NGC\,N4293. The nuclei of these three are indeed known to display luminous radio cores in high resolution 15\,GHz VLA images \citep{Nagar2005}. It is also seen that the red nuclei in the four Messier AGN host galaxies stand out, but we note that the red core emission in M\,87 must be attributed to strong non-thermal (synchrotron) radiation. We subsequently explore this infrared color selection for the extensive S$^{4}$G galaxy sample. 
\begin{table}
\caption[]{Properties of a few well-known nearby active galaxies.}
\begin{center}
\scalebox{0.80}
  {
\label{tab:nagn}
\centering
\begin{tabular}{lrcrlcc}
\hline\hline\noalign{\smallskip}
\multicolumn{1}{c}{Source} & \multicolumn{1}{c}{AGN} & \multicolumn{1}{c}{Hubble} &\multicolumn{1}{c}{$D$} & log($L_r$) & log($L_x$)  & $\log{M_{\rm{BH}}}$  \\
&\multicolumn{1}{c}{} &\multicolumn{1}{c}{} & \multicolumn{1}{c}{Mpc} & $\mathrm{erg}~\mathrm{s}^{-1}$& \multicolumn{1}{c}{$\mathrm{erg}~\mathrm{s}^{-1}$}&  $M_{\odot}$ \\
\multicolumn{1}{c}{(1)}&\multicolumn{1}{c}{(2)}&\multicolumn{1}{c}{(3)}&\multicolumn{1}{c}{(4)}&\multicolumn{1}{c}{(5)}&\multicolumn{1}{c}{(6)}&\multicolumn{1}{c}{(7)}\\

\noalign{\smallskip}\hline\noalign{\smallskip}
M\,81/NGC\,3031	& S1.5	& Sab	& 3.6	& 37.14		& 39.38	 & 7.76   \\
M\,87/NGC\,4486	& L2	& E0	& 16.8	& 39.03		& 40.78	 & 8.83   \\
M\,104/NGC\,4594 & L2	& Sa	& 9.2	& 37.90$^a$	& 40.69	 & 8.46   \\
M\,51/NGC\,5194	& S2	& Scd	& 7.7	& 35.50$^b$	& 41.03	 & 6.85   \\
\noalign{\smallskip}\hline
\end{tabular}
}
\tablefoot{\textit{Column (2)} AGN type from \cite{Ho1997}, \textit{Column (3)} Hubble type from NED, \textit{Column (4)} Distance, \textit{Columns (5-6)} Logarithm of radio (15~GHz; \citealt{Nagar2005}, 5~GHz; \citealt{Wang2003} ($^a$), 5~GHz; \citealt{Crane1992} ($^b$)) and X-ray luminosity from \cite{Ho2009},  \textit{Column (7)} Logarithm of black hole mass calculated from the velocity dispersion $\sigma$ using the $M_{\mathrm{BH}}-\sigma$ correlation from \cite{Tremaine2002}.}
\end{center}
\end{table}
\section{DATA AND SAMPLE CONSTRUCTION}
\label{sec:3}
In this study we use the \textit{Spitzer Survey of Stellar Structures in Galaxies} \citep[$\sfg$,][]{Sheth} sample of 2352 nearby galaxies, for which deep [3.6] and [4.5] micron images have been reduced and archived in collaboration with the DAGAL EU ITN network (Detailed Anatomy of GALaxies)\footnote{\url{http://www.dagalnetwork.eu/}}, supplemented with the $\sfg$ Extension\footnote{\url{http://sha.ipac.caltech.edu/applications/Spitzer/SHA/}}  sample of 464 galaxies. The $\sfg$ is a volume, magnitude, and size-limited survey of nearby galaxies, with deep Infrared Array Camera \citep[IRAC,][]{fazio} imaging at 3.6 and 4.5 \micron; it is one of the \textit{Spitzer} Legacy Programs of late type galaxies. This survey contains galaxies within 40 Mpc (v < 3000 km/s), away from the galactic plane (|b| > \(30^{\circ}\)), with extinction corrected  B-band magnitude brighter than 15.5 and B-band diameter larger than 1\arcmin\ with distances determined from HI radial velocities. Note that the fact that here HI velocities are required means that many early type galaxies are excluded, and for this reason the $\sfg$ Extension was observed.

We now apply the hot core method to the $\sfg$ galaxies with the aim to detect AGN by studying the central [3.6]-[4.5] color. We examined 2816 galaxies and obtained the central [3.6]-[4.5] color for 2741 galaxies in total. The nuclear [3.6] and [4.5] magnitudes were measured with an aperture of 3 pixels. The \textit{Spitzer} colors are measured using the IRAF\footnote{\url{http://iraf.noao.edu/}} \textit{apphot} package. The \textit{Spitzer} IRAC pixel size is 1.22\arcsec\ for all bands, yielding a measurement aperture of 3.66\arcsec\ in both bands. In this way we expect to be able to find AGN in a uniform way, without suffering from dust attenuation. The candidate AGN comprise galaxies of all Hubble types, allowing us to investigate the dependence of the AGN fraction upon Hubble type. We subsequently investigate whether these candidate AGN have radio detections. We use catalogues of large area radio surveys to check for the presence of radio counterparts. 

In addition to \textit{Spitzer} data, we also use data from the Wide-field Infrared Survey Explorer (\textit{WISE}). \textit{WISE} has mapped the whole sky in four infrared bands W1, W2, W3, and W4 centered at 3.4, 4.6, 12 and 22 \micron. We use \textit{WISE} to obtain two extra colors for the $\sfg$ galaxies, namely [W1-W2] and [W2-W3]. Band W4 is not used, given the large PSF. The \textit{WISE} pixel size is 1.37\arcsec\ for all bands, giving a minimum aperture of  5.48\arcsec\ . The central colors of the galaxies are measured for \textit{WISE}. We downloaded \textit{WISE} images of the $\sfg$ galaxies for these three bands from the \textit{WISE} Image Service \footnote{\url{http://irsa.ipac.caltech.edu/applications/wise/}}. This analysis is similar to the \textit{Spitzer} analysis, but has a factor $\sim{3}$ lower spatial resolution. Its results are presented in Appendix A. 

In the next sections we study the AGN fraction in the $\sfg$ survey, using the \textit{Spitzer} IRAC images. We then examine the fraction of AGN detected in this way that have detections in the radio from VLA FIRST\footnote{\url{ http://sundog.stsci.edu/first/catalogs.html}} and NVSS\footnote{\url{http://heasarc.gsfc.nasa.gov/W3Browse/radio-catalog/nvss.html}}. Finally, we also investigate the dependence of the AGN fraction on morphological types from the Third Reference Catalogue of Bright Galaxies \citep{deVaucouleurs} and the HyperLeda database\footnote{\url{http://leda.univ-lyon1.fr/}}.

\section{ANALYSIS}
\label{sec:4}
We determine the central colors for all $\sfg$ galaxies, and obtain candidate AGN by using the nuclear \textit{Spitzer} IRAC [3.6]-[4.5] colors. When doing the photometry, we eliminated saturated, shapeless (irregular) galaxies, and additionally a few galaxies for which the field was dominated by foreground stars. We analysed IRAC images of 2816 galaxies in total and eliminated 75 of them, yielding a final sample of 2741 galaxies. 
\subsection{Fractions of AGNs in the $\sfg$ sample, and their radio properties}
\label{subsec:4.1}

The near to mid-infrared regime provides a very good window to determine stellar masses, as the light emitted at these wavelenghts is dominated by K and M type giant stars, tracing the older stellar populations. As extinction is a strong function of wavelength, any stellar light emitted in the mid-infrared is significantly less affected by dust extinction than in the optical or shorter wavelenghts. \citet{Querejeta} used $\sfg$ galaxies and revealed the old stellar flux at  3.6 \micron\ and obtained stellar mass maps for more than 1600 galaxies in their work. They also separated the dominate light from old stars and the dust emission that can significantly contribute to the observed 3.6 \micron\ flux using the method described by \citet{Meidt2014} before. 

Emission from PAHs and hot dust can significantly  contribute to the flux detected at 3.6 \micron\ and 4.5 \micron\, in particular from the 3.3 \micron\ PAH feature and hot dust arising around active galactic nuclei \citep{Meidt2012a}. In the absence of PAH continuum, the [3.6]-[4.5] color is generally as low as [3.6]-[4.5] $\leq$ 0.15 (\citet{Querejeta}, \citet{Peletier2012,Norris2014}). \citet{Querejeta} modeled the color of a galaxy consisting of a stellar component, together with a component of hot dust (and PAHs). They conclude that the fraction of the emission in the [3.6] bands is not realistically larger than 0.4 in the $\sfg$ sample, which then implies that galaxies without AGN, even starburst galaxies, rarely reach [3.6]-[4.5] colors as red as 0.3. 

In this paper we apply a stringent infrared color criterion: we only consider those galaxies with central colors [3.6]-[4.5] $\geq{0.5}$. We subsequently investigate which of the AGN candidates have also been detected in the radio surveys. Examining the AGN detection rates as a function of morphological type, we classify these types into three main groups: early type galaxies (T-type -6, -5, -4, -3, -2, -1, 0), early type spirals (T-type 1, 2, 3, 4, 5) and late type spirals (T-type 6, 7, 8, 9, 10).

In Table~\ref{tab:percentages of AGNs}, we list the percentages and numbers of AGN (defined as having central [3.6]-[4.5] $\geq{0.5}$) and non-AGN ([3.6]-[4.5]< 0.5) according to morphological types. 1.3\% of the $\sfg$ galaxies appear to have [3.6]-[4.5] colors redder than 0.5 mag. This fraction is 0.8\% for early types, 2.4\% for early type spiral galaxies and 0.5\% for late type spiral galaxies. We also list the fractions of non-AGNs.

The 36 AGN are listed in Table B.1 in the Appendix; well-known Seyfert galaxies NGC\,1068, NGC\,4052, and NGC\,4151 are clearly detected. The nuclear mid-infrared color [3.6]-[4.5] is seen to range from 0.5 up to 1.6.  Interesting from Table~\ref{tab:percentages of AGNs} is that the AGN detection fraction is more prominent in early type spirals than in late-type spirals or early-type galaxies. This is entirely cf. expectation \citep[e.g.][]{Heckman2014}: ellipticals (like M~87) host AGN of the inefficient accretion type (FR1, or radio mode class); these objects lack dusty tori hence have little or no dust to reradiate the accretion disk luminosity. The spirals on the other hand will develop low luminosity (low mass black holes) AGN of efficient accretion (Seyfert-like) which do have dusty tori. Since early-type spirals have more massive bulges than late type spirals, the former can produce more AGN power. Most important: the numbers of AGN are small, implying that LINERS and other weak AGN types are not detected in this way.
\begin{table}
\caption[]{AGN fractions for the whole sample and the various morphological types separately.}
\label{tab:percentages of AGNs}
\begin{center}
\scalebox{0.98}
  {
\centering
\begin{tabular}{lcc}
\hline\hline\noalign{\smallskip}
\multicolumn{1}{l}{{\bf AGN:} type} & \multicolumn{1}{c}{{Fractions}}  & \multicolumn{1}{c}{{AGN (Total)}}  \\
\hline\noalign{\smallskip}
All             & 1.3 $\pm$ 0.2 \%       & 36 out of 2741 \\
Early type galaxies     & 0.8 $\pm$ 0.3 \%     &  6/728 \\
Early type spiral & 2.4 $\pm$ 0.5 \%     &  25/1053 \\
Late type spiral   & 0.5 $\pm$ 0.2 \%    &  5/957\\
\hline\hline\noalign{\smallskip}
\multicolumn{1}{l}{{\bf Non-AGN:} type}& \multicolumn{1}{c}{{Fractions}} & \multicolumn{1}{c}{{Non-AGN (Total)}}  \\
\hline\noalign{\smallskip}
All                      & 98.7 $\pm$ 0.2 \%  &  2705 out of 2741  \\
Early type galaxies      & 99.2 $\pm$ 0.3 \%     &  722/728 \\
Early type spiral       & 97.6  $\pm$ 0.5 \%     &   1028/1053 \\
Late type spiral        & 99.5  $\pm$ 0.2 \%    &   952/957\\
\hline\noalign{\smallskip}
\end{tabular}
}
\end{center}
\tablefoot{The AGN have been defined as those objects with central ([3.6]-[4.5] $>$ 0.5). We eliminated 3 galaxies lacking a TT type.
}
\end{table}

We now investigate whether these objects have radio continuum counterparts. The presence of a compact radio core in particular is a good indicator of the presence of an AGN, since its radio emission is mostly from non-thermal origin, i.e. not related to star formation (Ho 1999). Radio surveys are sensitive, and generally have very accurate positional information. We examine radio data from the low resolution -- 45\arcsec -- VLA NVSS, and the high resolution -- 5\arcsec -- VLA FIRST survey. If no radio flux for an infrared-selected candidate AGN is known, it does not necessarily mean that its radio flux is too low to be detected, but it is also possible that it has not been observed, given the limited sky coverage of the relevant radio survey. NVSS covered the sky north of $-40^o$, whereas the FIRST sky extends from $-10^o$ to $+65^o$. The former encompasses a large part of the $\sfg$: 2514 of the 2741 galaxies are within the NVSS sky; for FIRST that number is 1711 out of 2741. Table~3 lists the numbers of $\sfg$ sample galaxies with detections in the combined and in the two separate radio surveys. For NVSS we used a search radius of 45\arcsec, for FIRST 35\arcsec. There is obviously considerable overlap: most FIRST sources also appear in the NVSS catalog, but the reverse is not true.

\begin{table}
\caption[]{Fraction of the $\sfg$ galaxies detected in 2 radio surveys.}
\label{tab:S4G Radio detected}
\begin{center}
\scalebox{0.98}
  {
\centering
\begin{tabular}{lrc}
\hline\hline\noalign{\smallskip}
Surveys  & \multicolumn{1}{c}{Fractions} & \multicolumn{1}{c}{Detected ($\sfg$)} \\
\hline\noalign{\smallskip}
Radio     & 42.0 $\pm$ 1.0 \% & 1056 out of 2514 \\       
\hline\noalign{\smallskip}
VLA FIRST           & 20.3 $\pm$ 1.0 \%   &  347/1711 \\
NVSS             & 41.2 $\pm$ 1.0   \%   &  1037/2514 \\
\hline\noalign{\smallskip}
\end{tabular}
}
\end{center}
\end{table}
\begin{table}
\caption[]{Fractions and numbers of hot- and cold-core galaxies detected in the two radio surveys.}
\label{tab:S4G Hot-cold core}
\begin{center}
\scalebox{0.98}
  {
\centering
\begin{tabular}{lrc}
\hline\hline\noalign{\smallskip}
Surveys  & \multicolumn{1}{c}{Fractions} & \multicolumn{1}{c}{Hot-core ($\sfg$)} \\
\hline\noalign{\smallskip}
Radio     & 87.9 $\pm$ 5.7 \% &  29 out of 33 \\       
\hline\noalign{\smallskip}
VLA FIRST           & 80.0 $\pm$ 8.9 \%   &  16/20 \\
NVSS             & 87.9 $\pm$ 5.7   \%   &  29/33 \\
\hline\hline\noalign{\smallskip}
Surveys  & \multicolumn{1}{c}{Fractions} & \multicolumn{1}{c}{Cold-core ($\sfg$)} \\
\hline\noalign{\smallskip}
Radio     & 41.4 $\pm$ 1.0 \% & 1027 out of 2481 \\       
\hline\noalign{\smallskip}
VLA FIRST           & 19.6 $\pm$ 1.0 \%   &  331/1691 \\
NVSS             &  40.6 $\pm$ 1.0   \%   &  1008/2481 \\
\hline\noalign{\smallskip}
\end{tabular}
}
\end{center}
\tablefoot{The hot-core galaxies are the Spitzer-detected AGN detected in the radio, while the cold-core galaxies are only detected in the radio and not as infrared-detected AGN.
}
\end{table}
 \begin{figure}
    \centering    
    \resizebox{0.97\hsize}{!}{\includegraphics[]{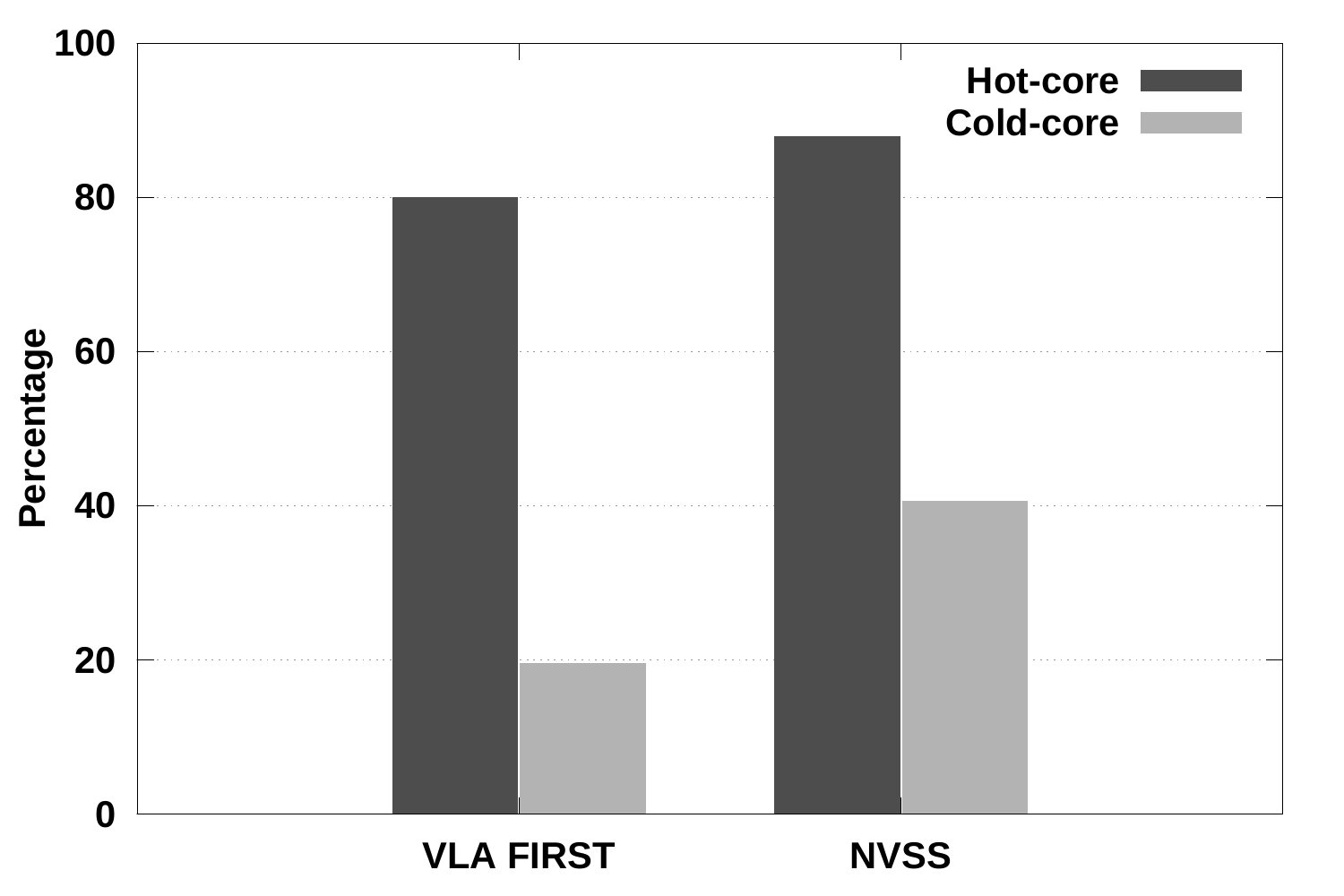}}
      \caption{The fraction of \textit{Spitzer} detected candidate AGN ([3.6]-[4.5] $\geq{0.5}$) also detected in other surveys for VLA FIRST and NVSS  (in dark colors in the histogram). Light colors in the histogram refer to the candidate non-AGN ([3.6]-[4.5] < {0.5}).}
    \label{fig:statistic1_figure}    
\end{figure}

Table~\ref{tab:S4G Radio detected} indicates that 42\% of the [3.6]-[4.5] color measured $\sfg$ galaxies are detected in radio surveys. We separate these galaxies as hot-core (candidate AGN) and cold-core (candidate non-AGN) in Table~\ref{tab:S4G Hot-cold core}. It is seen that 87.9\% of the hot-core galaxies and 41.4\% of the cold-core galaxies are detected in the radio surveys. Most interestingly is the difference between the low resolution NVSS and the high resolution FIRST survey. Whereas both detect comparable (high) fractions of candidate AGN, NVSS appears to resolve out many diffuse candidate non-AGN, confirming their proposed classification. These differing fractions for the two individual radio surveys are illustrated in the histograms of Figure~\ref{fig:statistic1_figure}. This figure also shows that the fractions of AGN that we detect in radio are very high, but significantly below 100\%, suggesting that our hot core method also finds radio-quiet AGN -- issues which require further study. The tables show that about 19\% of the infrared-detected candidate AGN have never before been detected as AGN in radio surveys. Also with reference to NGC\,4235 in Fig.~\ref{agnind}, we note that a less stringent red color cut, for instance 0.3 instead of 0.5, will undoubtedly yield more AGN candidates. 

Table~\ref{tab:detected_AGNs} in the Appendix lists the various hot core AGN, detected and/or undetected in the radio bands. Infrared detected candidate AGN which remain undetected in radio surveys are shown in bold-face font.
\subsection{Relation between galaxy luminosity and the presence of an AGN.}  
\label{subsec:4.2}
We now study the question whether the presence of an AGN depends on the absolute magnitude of the host galaxy. The absolute magnitude can be seen as a proxy of the galaxy stellar mass, which should correlate with the dynamical mass (through e.g. scaling relations like the Tully-Fisher relation).

\begin{figure}
    \centering
    \resizebox{0.97\hsize}{!}{\includegraphics[]{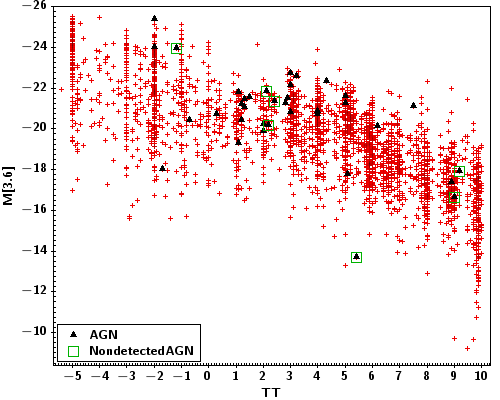}}
     \caption{Absolute magnitudes at 3.6 \micron\ ($M_{[3.6]}$) for the $\sfg$ sample galaxies versus morphological types (TT). Black triangle symbols refer to 36 AGN with [3.6]-[4.5] $\geq{0.5}$ and green big square symbols refer to AGN defined from their central colors [3.6]-[4.5] $\geq{0.5}$ not detected in radio.}
    \label{fig:MagTT_figure}
\end{figure}

In Figure~\ref{fig:MagTT_figure}, we plot the 2271 $\sfg$ sample galaxies with absolute magnitudes at 3.6 \micron\  (\citet{S4GCatalogue2011}) as a function of their optical morphological type. We now examine these detection rates as a function of morphological type. We have marked the 36 hot-core AGN as well as the radio undetected among these. For galaxies without absolute magnitudes at 3.6 \micron\, we used the Ks magnitudes from the 2MASS All Sky Extended Catalog (XSC)\footnote{\url{https://irsa.ipac.caltech.edu/Missions/2mass.html}}, converted to absolute magnitude using the distance modules  from the  NED\footnote{\url{http://ned.ipac.caltech.edu/}} database applying an average color K-[3.6] of 0.0 \citep[see][]{Falcon2011}. 

\begin{figure}
  \centering
   \resizebox{0.97\hsize}{!}{\includegraphics[]{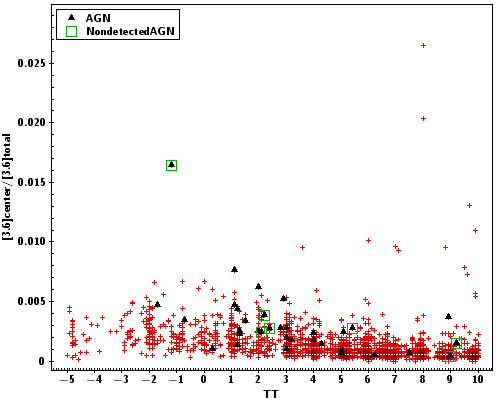}}
    \caption{Fraction of the light of the central source as a function of morphological type (TT). Black triangle symbols refer to 34 AGN with [3.6]-[4.5] $\geq{0.5}$ and green big square symbols refer to AGN defined from their central colors [3.6]-[4.5] $\geq{0.5}$ not detected in radio.}
    \label{fig:MagCenterTT}
\end{figure}
When compared to the whole sample, the 36 AGN in general seem to be  bright compared to the rest of the galaxies, and the majority of the AGN are spirals (as mentioned earlier). We find that the AGN are detected in the galaxies with the largest magnitudes/stellar masses. This is expected if the ratio of the red nucleus to the total galaxy luminosity is constant, since in this case the fainter AGN will not be detected. To test this we also plotted this ratio against morphological type. In Figure~\ref{fig:MagCenterTT} we show 2271 $\sfg$ sample galaxies with the relative luminosity at 3.6 $\mu$m of the central source as a function of morphological type. We have marked the 34 hot-core AGN as well as the radio undetected among these. We see that this relative luminosity is rather constant. Although there is a tendency that the AGN are brighter in the NIR for early-type spirals, also fainter AGN are detected, indicating that there is probably no strong type-dependent bias playing a role here. As mentioned in Section~\ref{subsec:4.1}, the relative occurrence of nuclear accretion and its dust re-radiation is cf. expectation.

\subsection{Confirmation with WISE data}
\label{subsec:4.3}

In addition to \textit{Spitzer} colors, we also obtained \textit{WISE} colors for our sample galaxies. The results of the comparison are shown in Appendix~\ref{AppA}. It is seen that -- despite the poorer \textit{WISE} resolution -- all \textit{Spitzer}-identified AGN stand out with a red [W1]$-$[W2] color (Fig. A.1). We subsequently applied the \citet{Mingo} \textit{WISE} color classification to indicate the host galaxy type of the $\sfg$ sample, both the AGN candidates and the non-AGN, in Fig. A.2. The equivalence of our \textit{Spitzer} AGN selection and the \citet{Mingo} \textit{WISE} AGN classification is clearly seen. Moreover,  \textit{Spitzer} detects AGNs which the \textit{WISE} diagnostic describes as ULIRG/obscured AGN. Again -- like above -- dusty spiral galaxies dominate as hosts for these low-luminosity AGNs.

\section{Discussion and conclusions}
\label{sec:5}
We have investigated the central \textit{Spitzer} [3.6]-[4.5] colors of 2741 galaxies of the $\sfg$ and the $\sfg$ extension sample to study the presence of AGN in nearby galaxies using a method inspired by \citet{stern2005}. We conclude that galaxies with central colors [3.6]-[4.5] $\geq{0.5}$ are likely to be AGN. They have very red cores, which we attribute to AGN heated circumnuclear dust.

We compared our results with the literature. As mentioned earlier, summed over all Hubble types 40\% of the galaxies can be considered active \citep{Ho2008}. The nearby galaxy survey of \citet{kauffmann&heckman} examine the properties of the host galaxies of 22 623 narrow-line, obscured AGN with $0.02 < z < 0.3$ selected from a complete sample of 122 808 galaxies from the SLOAN Digital Sky Survey (SDSS; \citet{York2000}; \citet{Stoughton2002}). \citet{kauffmann&heckman} report an overall AGN fraction  of $\sim$20\%, of which $\sim$10\% are Seyferts, based on the optical survey of the SDSS. They focus on the luminosity of the [\ion{O}{III}] $\lambda{5007}$ emission line as a tracer of the strength of activity in nucleus. They show that the majority of AGN in their sample fall into the ‘transition’ class, having line ratios intermediate between those of star-forming galaxies and those of LINERs (low-ionization nuclear emission-line regions) or Seyferts. They show that low-luminosity AGN have old stellar populations similar to those of early-type galaxies. High-luminosity AGN reside in significantly younger hosts, that is in ordinary late-type galaxies. In our study, measuring the color excess of a sample of S4G galaxies, we find an AGN fraction of 1.3\%.  42\% of $\sfg$ galaxies are detected in radio surveys and 88\% of those are also detected in radio surveys as a hot-core (AGN candidate). Our detection statistics are not as high as optical spectroscopic surveys, but using the near-infrared selection method with radio comparison is also effective to find AGN. Besides that, our AGN detection fraction is more prominent in early type spirals than in late-type spirals or early-type galaxies. We here find AGN which have infrared-bright tori, and most of these mid-IR selected AGN have early type spiral host galaxies. Our survey obviously fails to detect AGN in early type hosts, as these lack dust.

Recently \citet{Kauffmann2018} published a study of AGN, selected using the infrared method described in this paper. They examine the narrow emission-line properties and stellar populations of a sample of 1385 radio-detected, mid-IR excess AGNs in order to understand the physical conditions in the interstellar medium of these objects. They compare these systems with a control sample of 50 000 AGNs selected by their optical emission-line ratios that do not have a significant mid-IR excess. Starting with a sample of SDSS (with redshifts up to 0.5), and combining them with the AllWISE survey \citep{Wright2010} this author finds that, although most of the optically defined AGN are blue in the [W1]-[W2] color, there is an extended tail towards very red [W1]-[W2] colors, i.e., AGN that we define here. The author reports a surprisingly large fraction of normal galaxies with very red central colors, as we do in this paper. Although most of the objects are too far away to resolve spatially, \citet{Kauffmann2018} finds, for the nearest galaxies, that radio luminosity is the quantity that is most predictive of a redder central [W1]-[W2] color. Of the radio-detected AGN, 80\% would also be optically classified as an AGN. \citet{Kauffmann2018} concludes that the more detailed studies support the view that galaxies with centrally peaked mid-IR emission are those where black hole growth may be occurring in a mode that is largely hidden at optical wavelengths, and that black hole growth may be modulated/regulated by energetic feedback from relativistic jets generated by the accreting black hole. Our paper is complementary to this paper, the AGN statistics (88\%) determined by the near-infrared color excess method with radio survey counterpart detection used in our work is higher, since it is studying the statistical properties of a nearby sample, using the higher spatial resolution Spitzer data, confirming some of the main results of \citet{Kauffmann2018}.

The nearby active nuclei show a near-infrared excess in comparison to normal nuclei which are dominated by evolved stellar and star formation emission. This gives rise to the interpretation that this excess is connected to supermassive black hole accretion. The near-infrared emission of active nuclei can be described as arising from either hot circumnuclear dust or a non-thermal power-law spectral distribution. The sources with this excess all have compact radio cores with high brightness temperature \citep{vanderwolk2011}. The mid-infrared emission shows that these hot dust components are often surrounded by warm dust tori, these are particularly found in luminous radio-loud AGN \citep{Meisenheimer2001, Ogle2006}. In our work, we find the fraction of AGN with a strong near-infrared [3.6]-[4.5] color excess. 88\% of these AGN are detected in the archives of radio surveys, respectively. This reveals that these AGN have compact radio cores and also have bright infrared obscuring torus.

We pre-classified our sources based on their mid-IR colors, using the \textit{WISE} color/color plot and the classification of \citet{Mingo}, as elliptical, spiral, starburst and AGN. The sources we found as AGN with \textit{Spitzer} were almost all confirmed using \textit{WISE} data. 

Star formation histories of spiral galaxies show that star formation occurs as circumnuclear starbursts in the spiral galaxies, with the highest star formation rates in early type spiral galaxies, and possibly related to the presence of bar and often co-occurs with an AGN \citep{Kormendy2004,Knapen2006}. The existence of almost bulge-less, late spiral galaxies provides a clue that undisturbed galaxies exist. AGN, for which matter needs to flow into the centre, are uncommon in bulgeless galaxies, and common in galaxies with bulges \citep{Ho2008}. Apparently, as spiral galaxies get older their bulges and nuclear supermassive black holes are increasingly fed by material from the disc. In our work, we find that the fraction of IR-defined AGN is highest for early type spiral galaxies in the nearby universe, in apparent agreement with the high star formation rates for such galaxies. 

We showed that the \textit{Spitzer Space  Telescope}  is a efficient, powerful tool  for  studying  AGN demographics.  The color-color method represents an efficient way of distinguishing AGN from normal galaxies and an efficient way of finding AGN in a large sample. These diagnostics can prove to be extremely useful for large AGN and galaxy samples, and help develop ways to efficiently classify objects when data from the next generation of instruments become available.

\begin{acknowledgements}
Tuba \.{I}kiz acknowledges a PhD scholarship from The Scientific and Technological Research Council of Turkey (T\"{U}B\.{I}TAK) under project number 1059B141400919 and 1649B031406125 and is supported by the Atat\"{u}rk University. This publication is based on observations obtained with the \textit{Spitzer Space Telescope}, which is operated by the Jet Propulsion Laboratuary, California Institute of Technology, under a contract with the National Aeronautics and Space Administration (NASA) and the \textit{Wide-field Infrared Survey Explorer}, which is a joint project of the University of California, Los Angeles, and the Jet Propulsion Laboratory/ California Institute of Technology, funded by the National Aeronautics and Space Administration (NASA). We wish to thank Guido van der Wolk for contributions to improve the paper. We would like to thank the DAGAL team for their great work to make $\sfg$ sample data public and available. We would like to thank the hospitality of Kapteyn Astronomical Institute. This research has made use of the NASA/IPAC Extragalactic Database (NED) which is operated by JPL, Caltech, under contract with NASA. We acknowledge the use of the FIRST and NVSS catalogues, provided by NRAO. We thank the anonymous referee for insightful and constructive comments that greatly improved the paper. We would like to thank the SUNDIAL ITN network for support. RFP  acknowledges financial support from  the  European Union's Horizon 2020 research and innovation program under the Marie  Sklodowska-Curie grant  agreement  No.  721463  to  the SUNDIAL ITN network We also thank the TOPCAT developer Mark Taylor for programming and releasing a software patch so that we could improve our figures. 
\end{acknowledgements}

\bibliographystyle{aa}
\bibliography{tuba.bib}

\begin{appendix}

\section{\textbf{\textit{Spitzer} and \textit{WISE} color comparison}}
\label{AppA}

\begin{figure*}
    \centering
\resizebox{0.97\hsize}{!}{\includegraphics[]{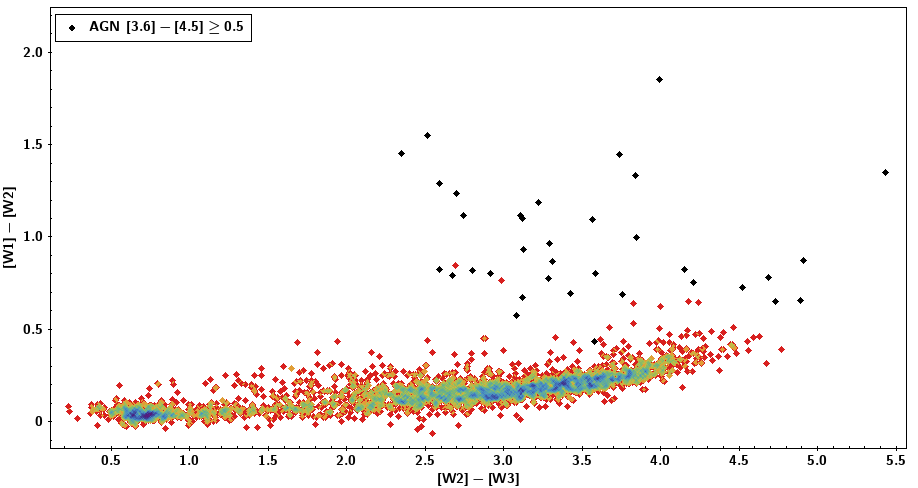}}
    \caption{\textit{WISE} [W1]-[W2] color versus [W2]-[W3] color magnitude density diagram for the $\sfg$ sample. Black symbols refer to \textit{Spitzer} detected AGN [3.6]-[4.5] $\ge{0.5}$.}
    \label{fig:spitzer_wise_figure}    
\end{figure*}

\begin{figure*}
    \centering
\resizebox{0.97\hsize}{!}{\includegraphics[]{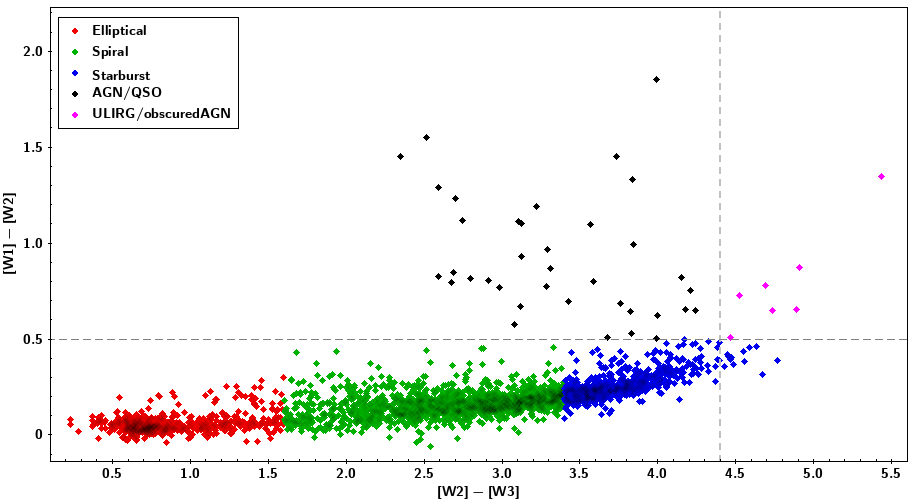}}
    \caption{[W1]-[W2] versus [W3]-[W3] color - color magnitude diagram for the galaxies in $\sfg$ sample. The symbols are based on the \citet{Mingo} classification of Table~\ref{tab:Activity_Table}, whereby ULIRG/obscured AGN with ([W1]-[W2] > 0.5, [W3]-[W3] $\ge {4.4}$) are indicated with pink symbols.}
    \label{fig:wise_wise_cc_figure}
\end{figure*}

In this Appendix we show the \textit{WISE} [W1]-[W2] versus [W2]-[W3] color-color plot in Figure~\ref{fig:spitzer_wise_figure} and the activity diagnostics in Figure~\ref{fig:wise_wise_cc_figure} based on \cite{Mingo} classification in Table~\ref{tab:Activity_Table}, for our $\sfg$ galaxy sample.
The equivalence of our \textit{spitzer} hot core selection and the \citet{Mingo} AGN classification using \textit{WISE} [W1]-[W2] color is clearly demonstrated. Note that IC\,0630 is an AGN according to its \textit{WISE} data, but is missing in the \textit{Spitzer} AGN list due to saturation. Reversely, NGC\,0253 is a \textit{Spitzer} AGN, but not a \textit{WISE} AGN, due to saturation.

We pre-classified the $\sfg$ sources using the scheme of  \citet{Mingo}; the resulting color/color diagram is shown in  Figure~\ref{fig:wise_wise_cc_figure}. We used Table~\ref{tab:Activity_Table}, taken from \citet{Mingo}, which describes in detail the class boundaries (see e.g. the source distributions on the equivalent WISE color-color plots of \citet{Gurkan_2014,Yang_2015}). We classify our sources using these boundaries and also show the sources which are not described by \citet{Mingo} with pink symbols. The latter group refers to ULIRG/obscured AGNs ([W1]-[W2] > 0.5, [W3]-[W3] $\ge {4.4}$). All sources we find as an AGN candidate on the basis of their \textit{Spitzer} color [3.6]-[4.5] $>$ 0.5 mag can also be classified as such on the basis of their \textit{WISE} color.
\begin{table*}\small
\caption{Activity Table based on \citet{Mingo}. For each of our source types, selected on the \textit{WISE} color/color plot, this table shows the types of activity most likely to be found at each wavelength. Please note that, for each color category, several combinations of the elements in columns 2-4 may be possible, e.g. in the first group, an elliptical galaxy in a cluster, with a radiatively inefficient AGN in X-rays, and an LERG in radio. LINER stands for low-ionization nuclear emission-line region. ULIRG stands for ultraluminous infrared galaxy. QSO stands for quasars. LERG stands for low excitation radio galaxy; high excitation sources (HERG) include NLRG (narrow-line radio galaxies) and BLRG (broad line radio galaxies).}
\label{tab:Activity_Table}
\centering
\begin{tabular}{lllll}
\hline\hline\noalign{\smallskip}
Label & WISE color selection & Mid-IR/Optical & X-rays & Radio \\
\hline\noalign{\smallskip}
\multirow{3}{*}{Elliptical} & \multirow{3}{*}{$[W1]-[W2]<0.5$;  $0<[W2]-[W3]<1.6$} & Elliptical galaxy (isolated) & \multirow{2}{*}{Rad. inefficient AGN} & \multirow{3}{*}{LERG} \\
&& Elliptical galaxy (cluster) & \multirow{2}{*}{Hot ICM gas} & \\
&& LINER & \multirow{-1}{*}{}&\\\hline

\multirow{3}{*}{Spiral} & \multirow{3}{*}{$[W1]-[W2]<0.5$;  $1.6\leq [W2]-[W3]<3.4$} & \multirow{2}{*}{Star-forming galaxy} & \multirow{2}{*}{Star formation} & Star formation \\
&&\multirow{2}{*}{Star-forming galaxy + AGN} & \multirow{2}{*}{Seyfert galaxy} & Low-L NLRG\\
&&&& LERG \\
\hline\noalign{\smallskip}

\multirow{2}{*}{Starburst}&\multirow{2}{*}{$[W1]-[W2]<0.5$; $[W2]-[W3] \geq 3.4$}&Starburst galaxy&Star formation&Star formation\\
&&ULIRG&Seyfert galaxy&Low-L NLRG\\
\hline\noalign{\smallskip}

\multirow{3}{*}{AGN/QSO} & \multirow{3}{*}{$[W1]-[W2] \geq 0.5$; $[W2]-[W3]<4.4$} & \multirow{3}{*}{AGN} & Luminous Seyfert galaxy & NLRG \\
&&&BL-Lac&BLRG\\
&&&QSO&QSO\\
\hline\noalign{\smallskip}
\end{tabular}
\end{table*}
\section{\textit{Spitzer} and \textit{WISE} colors of the $\sfg$ sample}
\label{AppB}
In this Appendix, we present a list of the radio detected and non-detected AGN in related radio surveys (FIRST, NVSS) with [3.6]-[4.5] $\ge{0.5}$ mag for the $\sfg$ sample. For Table~\ref{tab:detected_AGNs} column 1 gives the galaxy names, column 2-3 contains their right ascension and declination in degrees at epoch J2000.0, and in columns 4 we indicate the numerical morphological type, and in column 5-10 we gives the the [3.6]-[4.5], [W1]-[W2], and [W3]-[W3] colors with their errors, and in column 11 we present  galaxies as 'detected' and 'not detected' in radio surveys, and in column 12 we indicate the radio surveys that galaxies detected or non-detected, respectively.
\begin{table*}
\caption[]{AGN defined using their central \textit{Spitzer} colors [3.6]-[4.5] $\geq{0.5}$.}
\label{tab:detected_AGNs}
\small
\centering
\begin{adjustbox}{width=1\textwidth}
\renewcommand{\arraystretch}{1.3}
\begin{tabular}{lrrrcccccccc}
\hline\hline\noalign{\smallskip}
\multicolumn{1}{c}{\textbf{Galaxy}} & \multicolumn{1}{c}{\textbf{RA}}    & \multicolumn{1}{c}{\textbf{DEC}}   & \textbf{TT}  & \textbf{{[}3.6{]}-{[}4.5{]}} & \textbf{Error } & \textbf{{[}W1{]}-{[}W2{]}} & \textbf{Error} & \textbf{{[}W2{]}-{[}W3{]}} & \textbf{Error}& \textbf{Radio} & \textbf{Survey}  \\[0.02cm]
 & \multicolumn{1}{c}{\textbf{(deg)}} & \multicolumn{1}{c}{\textbf{(deg)}} & \textbf{}    & \textbf{(mag)} & \textbf{(mag)}   & \textbf{(mag)} & \textbf{(mag)}  & \textbf{(mag)} & \textbf{(mag)}   & \textbf{(detected/non-detected)} & \textbf{(FIRST, NVSS)}          \\[0.02cm]
\multicolumn{1}{c}{\textbf{(1)}}    & \multicolumn{1}{c}{\textbf{(2)}}   & \multicolumn{1}{c}{\textbf{(3)}}   & \textbf{(4)} & \textbf{(5)}                & \textbf{(6)}   & \textbf{(7)}   & \textbf{(8)}    & \textbf{(9)}      & \textbf{(10)} & \textbf{(11)} & \textbf{(12)}      \\[0.02cm]
\hline\noalign{\smallskip} 
ESO 409-015 &	1.38364		&	-28.09991	&	5.4	&	0.800 	&	0.413	&	0.686	&	0.124  &	3.759		&	0.106 & \textbf{non-detected} &--\\[0.02cm]
NGC 0253	 &	11.86515	&	-31.42178	&	-1.2&	0.789 	&	0.021	&	-- 		&	 --	   &	 --			&	 --	  & \textbf{non-detected}  & -- 	\\[0.02cm]
NGC 0520	 &	21.14538	&	3.79159		&	1.3	&	0.522 	&	0.043	&	0.752	&	0.016  &	4.207		&	0.013 & detected & FIRST, NVSS	\\[0.02cm]
NGC 0625	 &	23.76455	&	-41.43722	&	9.0	&	0.688 	&	0.137	&	0.654	&	0.043  &	4.890		&	0.035 &  \textbf{non-detected} & -- \\[0.02cm]
NGC 0660	 &	25.75969	&	13.64581	&	1.3	&	0.568 	&	0.026	&	1.189	&	0.007  &	3.220		&	0.006 & detected & NVSS		\\[0.02cm]
NGC 0814	 &	32.65672	&	-15.77344	&	-1.7&	0.744 	&	0.108	&	0.780	&	0.039  &	4.688		&	0.031 & detected  & NVSS	\\[0.02cm]
PGC 009354	&	36.88651	&	-10.16587	& 5.1	& 0.843 &	0.199 & 0.871 &	0.068 & 4.912 & 0.053  & detected & NVSS	\\[0.02cm]
NGC 1068	 &	40.66962	&	-0.01331	&	3.0	&	0.797 	&	0.017	&	1.116	&	0.006  &	2.744		&	0.005 & detected	& FIRST, NVSS	\\[0.02cm]
NGC 1365	 &	53.40155	&	-36.14039	&	3.2	&	0.679 	&	0.020	&	0.800	&	0.006  &	3.587		&	0.005 & detected	& NVSS	\\[0.02cm]
IC 1953	&	53.42431	&	-21.47868	& 6.2	& 0.989 &	0.150 & 0.728 &	0.045 & 4.522 & 0.037 & detected &	NVSS \\[0.02cm]
NGC 1386	 &	54.19239	&	-35.99920	&	-0.7&	0.783 	&	0.033	&	0.965	&	0.011  &	3.289		&	0.009 & detected & NVSS \\[0.02cm]
NGC 3034	 &	148.96800	&	69.67975	&	7.5	&	0.690 	&	0.013	&	1.453	&	0.004  &	2.351		&	0.004 &  detected	& NVSS \\[0.02cm]
NGC 3094	&	150.35812	&	15.77011	& 1.1	& 0.950 &	0.028 & 1.549 &	0.008 & 2.513 & 0.006 & detected & FIRST, NVSS	\\[0.02cm]
NGC 3227	 &	155.87740	&	19.86513	&	1.5	&	0.590 	&	0.026	&	0.775	&	0.011  &	3.287		&	0.009 & detected & FIRST, NVSS	\\[0.02cm]
NGC 3516	 &	166.69780	&	72.56850	&	-2.0&	0.519 	&	0.035	&	0.825	&	0.009  &	2.593		&	0.009 & detected & NVSS	\\[0.02cm]
UGC 06433	&	171.38258	&	38.06064	& 9.2	& 0.834 &	0.308 & 0.651 &	0.100 & 4.733 & 0.081 & \textbf{non-detected}  &-- 	\\[0.02cm]
NGC 3729	 &	173.45578	&	53.12555	&	1.2	&	0.704 	&	0.067	&	0.693	&	0.025  &	3.424		&	0.021	& detected	 & FIRST, NVSS \\[0.02cm]
NGC 4051	 &	180.79007	&	44.53131	&	4.0	&	0.748 	&	0.030	&	0.931	&	0.018  &	3.126		&	0.015	& detected &	FIRST, NVSS \\[0.02cm]
NGC 4151	&	182.63561	&	39.40578	& 2.0	& 0.988 &	0.029 & 1.233 &	0.010 & 2.699 & 0.009 & detected	 & FIRST, NVSS	\\[0.02cm]
NGC 4293	 &	185.30347	&	18.38261	&	0.3	&	0.635 	&	0.048	&	0.672	&	0.032  &	3.121		&	0.029 & detected &	FIRST, NVSS	\\[0.02cm]
NGC 4388	&	186.44490	&	12.66209	& 2.8	& 0.855 &	0.033 & 1.114 &	0.020 & 3.108 & 0.016  & detected & FIRST, NVSS  	\\[0.02cm]
NGC 4355	&	186.72764	&	-0.87767	& 1.1	& 1.358 &	0.066 & 1.348 &	0.023 & 5.433 & 0.016 & detected & FIRST, NVSS	\\[0.02cm]
NGC 4593	 &	189.91437	&	-5.34414	&	3.0	&	0.667 	&	0.035	&	0.816	&	0.020  &	2.800		&	0.018 & detected & FIRST, NVSS	\\[0.02cm]
NGC 4628	&	190.60520	&	-6.97103	& 2.9	& 0.812 &	0.045 & 0.869 &	0.029 & 3.310 & 0.025 & detected &	FIRST, NVSS	\\[0.02cm]
ESO 443-042 &	195.87400	&	-29.82870	&	3.0	&	0.597 	&	0.091	&	0.573	&	0.031  &	3.081		&	0.029 & detected &	NVSS	\\[0.02cm]
NGC 4968	 &	196.77420	&	-23.67690	&	-2.0&	0.744 	&	0.045	&	0.995	&	0.018  &	3.844		&	0.014 & detected & NVSS	\\[0.02cm]
NGC 5253	&	204.98315	&	-31.64006	& 8.9	& 1.298 &	0.027 & 1.852 &	0.011 & 3.991 & 0.007 & detected	& NVSS	\\[0.02cm]
NGC 5347	&	208.32421	&	33.49085	& 2.0	& 0.890 &	0.044 & 1.095 &	0.031 & 3.566 & 0.025 & detected	& FIRST, NVSS	\\[0.02cm]
NGC 5427	 &	210.85854	&	-6.03075	&	5.0	&	0.506 	&	0.082	&	0.432	&	0.027  &	3.579		&	0.024 & detected  &	FIRST, NVSS \\[0.02cm]
NGC 5506	&	213.31209	&	-3.20757	& 1.2	& 1.059 &	0.029 & 1.292 &	0.011 & 2.591 & 0.009 & detected	& FIRST, NVSS	\\[0.02cm]
NGC 5861	&	227.31709	&	-11.32171	& 5.0	& 1.601 &	0.084 & 1.331 &	0.025 & 3.837 & 0.018  & detected  &   NVSS 		\\[0.02cm]
NGC 7314	 &	338.94252	&	-26.05043	&	4.0	&	0.712 	&	0.047	&	0.804	&	0.018  &	2.912	&	0.017 & detected & NVSS	\\[0.02cm]
NGC 7378	 &	341.94867	&	-11.81664	&	2.2	&	0.803 	&	0.072	&	0.793	&	0.027  &	2.673	&	0.026 & \textbf{non-detected}  &  --	\\[0.02cm]
NGC 7479	&	346.23590	&	12.32293	& 4.3	& 1.150 &	0.041 & 1.449 &	0.015 & 3.735 & 0.011 & detected	& FIRST, NVSS	\\[0.02cm]
NGC 7552	 &	349.04494	&	-42.58496	&	2.4	&	0.511 	&	0.028	&	0.822	&	0.008  &	4.154		&	0.006 & \textbf{non-detected} &	--	\\[0.02cm]
NGC 7582	&	349.59837	&	-42.37034	& 2.1	& 0.890 &	0.026 & 1.100 &	0.006 & 3.122 & 0.006 & \textbf{non-detected} &	--	\\[0.02cm]
\hline\noalign{\smallskip}
\end{tabular}
\end{adjustbox}
\tablefoot{\scriptsize Columns are: \textbf{(1)} Galaxy name; \textbf{(2)} Right ascension (J2000); \textbf{(3)} Declination (J2000); \textbf{(4)} Numerical morphological type; \textbf{(5)} [3.6]-[4.5] color; \textbf{(6)} [3.6]-[4.5] color error; \textbf{(7)} [W1]-[W2] color; \textbf{(8)} [W1]-[W2] color error; \textbf{(9)} [W2]-[W3] color; \textbf{(10)} [W2]-[W3] color error; \textbf{(11)} Radio (detected/non-detected) ; \textbf{(12)} Survey. Column \textbf{(11)} indicates that whether AGN detected or non-detected in both surveys. Column \textbf{(12)} refers that whether AGN ([3.6]-[4.5] $\geq{0.5}$) are detected in the archives of VLA FIRST and NVSS within 35\arcsec\ and 45\arcsec\ radius, respectively. (--) refers lack of data in both surveys. 
}
\end{table*}

\end{appendix}
\end{document}